\magnification = 1200
\baselineskip = 1.2\baselineskip

\input harvmac.tex

\def\oh{\overline{h}}
\def\og{\overline{g}}
\def\oD{\overline{D}}
\def\oA{\overline{A}}
\def\oPhi{\overline{\Phi}}
\def\hg{\widehat{g}}
\def\hs{\widehat{s}}
\def\vv{\vec{v}}
\def\vy{\vec{y}}

\hfuzz=5pt

\def\NP{Nucl.\ Phys.\ }

\lref\bfss{T. Banks, W. Fischler, S.H. Shenker and L. Susskind, {\it
Phys. Rev.} {\bf D55} (1997) 5112, {\tt hep-th/9610043}.}
\lref\mald{J. Maldacena, ``The large $N$ limit of superconformal field
theories and supergravity'', {\tt hep-th/9711200}.}
\lref\maldetal{N. Itzhaki, J. Maldacena, J. Sonnenschein and S. Yankielowicz,
``Supergravity and the large $N$ limit of theories with sixteen 
supercharges'',  {\tt hep-th/9802042}.}
\lref\dkps{M. Douglas, D. Kabat, P. Pouliot and S. Shenker,
\NP {\bf B485} (1997) 85, {\tt hep-th/9608024}.}
\lref\beco{D. Berenstein and R. Corrado, {\it Phys. Lett. B} {\bf 406}
(1997) 37, {\tt hep-th/9702108}.}
\lref\bebe{K. Becker, M. Becker, \NP {\bf B506} (1997) 48, 
{\tt hep-th/9705091}.}
\lref\bbpt{K. Becker, M. Becker, J. Polchinski and A. Tseytlin, {it
   Phys. Rev.} {\bf D56} (1997) 3174, {\tt hep-th/9706072}.}
\lref\seib{N. Seiberg, {\it Phys. Rev. Lett.} {\bf 79} (1997) 3577,
{\tt hep-th 9710009}.}
\lref\sen{A. Sen, ``D0 branes on $T^n$ and matrix theory'', {\tt
hep-th/9709220}.}
\lref\senrev{A. Sen, ``An introduction to nonperturbative string theory'', 
{\tt hep-th/9802051}.}
\lref\gukl{S. Gubser and I. Klebanov, Phys.Lett. {\bf 413B} (1997) 41,
{\tt hep-th/9708005}.}
\lref\hepo{S. Hellerman and J. Polchinski, ``Compactification in the
lightlike limit,'' {\tt hep-th/9711037}.}
\lref\bila{A. Bilal, ``DLCQ of M-theory as the light-like limit'', 
{\tt hep-th/9805070}.}
\lref\hyun{S. Hyun, ``The background geometry of DLCQ supergravity'',
{\tt hep-th/9802026}.}
\lref\hyunetal{S. Hyun, Y. Kiem and H. Shin, {\it Phys. Rev.} {\bf D57}
(1998) 4856, {\tt hep-th/9712021}.}
\lref\dob{K.S. Stelle, ``BPS branes in supergravity'', {\tt hep-th
9803116}.}
\lref\witt{E. Witten, Nucl.Phys. B443 (1995) 85, {\tt hep-th/9503124}.}
\lref\polchrr{J. Polchinski, Phys. Rev. Lett. {\bf 75} (1995) 4724,
{\tt hep-th/9510017}.}
\lref\eard{R. Ferrell and D. Eardley, Phys. Rev. Lett. {\bf 59} (1987) 1671.}
\lref\banks{T. Banks, Nucl. Phys. Proc. Suppl. {\bf 67} (1998) 180,  
{\tt hep-th/9710231}.}
\lref\dira{M. Dine and A. Rajaraman, ``Multigraviton scattering in
the matrix model'', {\tt hep-th/9710174}.}
\lref\susso{L. Susskind, ``Another conjecture about M(atrix) theory'', 
{\tt hep-th/9704080}.}
\lref\bigsuss{D. Bigatti and L. Susskind, ``Review of Matrix theory'', 
{\tt hep-th/9712072}.}
\lref\flato{C. Bachas, Phys.Lett. {\bf 374B} (1996) 37, {\tt hep-th/9511043}.}
\lref\flattw{R. Khuri and R. Myers, Nucl.Phys. {\bf B466} (1996) 60,
{\tt hep-th/9512061}.}
\lref\aj{J. McCarthy and A. Wilkins, to appear.}
\lref\ffi{M. Fabbrichesi, G. Ferretti and R. Iengo,
``Supergravity and Matrix theory do not disagree on multi-graviton
scattering'', {\tt hep-th/9806018}.}
\lref\wtr{W. Taylor IV and M. Raamsdonk, ``Three-graviton scattering in Matrix
theory revisited'', {\tt hep-th/9806066}.} 
\lref\ecgr{R. Echols and J. Gray, ``Comment on multigraviton scattering 
in the Matrix model'', {\tt hep-th/9806109}.}
\lref\okyo{Y. Okawa and T. Yoneya, ``Multi-body interactions of D-particles in 
supergravity and Matrix theory'', {\tt hep-th/9806108}.}

\line{}
\vskip2cm

\centerline{\titlefont Large $N$ and the Dine-Rajaraman problem}
\vskip2cm

\centerline{Jim McCarthy$^a$, 
Leonard Susskind$^b$
\foot{Supported by the NSF under Grant No. PHY-9219345.}
and Andy Wilkins$^a$}\bigskip

\centerline{\sl $^a$ Department of Physics and Mathematical Physics}
\centerline{\sl University of Adelaide, Adelaide, SA~5005, Australia}

\vskip1cm

\centerline{\sl $^b$ Department of Physics}
\centerline{\sl  Stanford University, Stanford, CA~94305-4060, USA}

\vskip1.5cm

\centerline{\bf Abstract}\smallskip

\noindent We compute the effective action for scattering of
three well-separated extremal brane solutions, in 11d supergravity,
with zero $p_-$ transfer and small transverse velocities.  Using an
interpretation of the conjecture of Maldacena, following Hyun, this
can be viewed as the large $N$ limit of the Matrix theory description
of three supergraviton scattering at leading order.  The result is
consistent with the perturbative supergravity calculation.

\vskip1.5cm

\vfil
\line{ADP-98-30/M68\hfil}
\line{{\tt hep-th/9806136}\hfil June 1998}

\eject


\newsec{Introduction}

 Matrix theory \bfss\ proposes that M-theory is described by the
maximally supersymmetric quantum mechanics of U($N$) matrices in the
large $N$ limit.  Moreover, a prescription for computations is given
\bfss \seib \sen\ in which it has been argued \susso\ that finite $N$ 
corresponds to the so-called discrete light cone frame quantization of
the M-theory (for a review see \senrev).  One test of these
conjectures (see, e.g., \bebe \bbpt) is to compare low energy scattering
of supergravitons in the Matrix theory with the corresponding results
in eleven-dimensional supergravity -- a subtle limit, see \hepo\bila.  In
particular, agreement is found for the scattering of two
well-separated supergravitons with small transverse velocities \dkps
\bfss \beco.  However, supergravity seems to predict different behaviour 
to that of the matrix model (at any finite order in $N$) for processes
such as the scattering of three supergravitons \dira.

 It has been suggested \banks \bigsuss\ that this discrepancy may
vanish on taking the large $N$ limit.  However, only through the
recent work of Maldacena \mald\ has it been possible to deal with this
limit.  In \mald \maldetal\ brane configurations were studied in the
limit where the field theory on the brane decouples from the bulk, and
it was observed that when the number of branes $N$, becomes large, the
curvature of spacetime around the brane becomes small (for earlier
discussions in the conformal case see \gukl\ and references therein).
However, for small curvatures branes are well described by extremal
black-hole type solutions of the associated supergravity.  Moreover,
as discussed in \hyun, this limit corresponds to the infinite boost
limit in the DLCQ Matrix theory prescription mentioned above.  Thus we
are naturally led to the following conjecture --- which we take to be
the premise of this work --- that in the large $N$ limit of DLCQ
Matrix theory, supergravitons are described by D0-brane solutions of
IIA supergravity.  Since these D0-branes are BPS states which can be
identified with Kaluza-Klein supergraviton modes of 11d supergravity
\witt\polchrr, this conjecture immediately implies a resolution of the 
problem in \dira.  The leading order scattering amplitudes will just
be proportional to those of point particles in 11d supergravity, as
are those of supergravitons.  In the rest of this paper we describe an
explicit calculation of the three supergraviton amplitude as
``extremal black hole'' solutions, since the details may be of
interest.

  Thereto, we calculate the effective action for large separation and
low transverse velocity scattering of these particles (neglecting spin
effects as usual), following a ``post-Newtonian'' calculation similar
to those in, e.g., \eard\ -- with the slight twist that we work in
the lightcone frame.  The essence of the calculation is that we
promote the centers in the static solution to dynamical variables, and
then determine corrections to the metric so that we have a solution
\eqn\jmoa{ g_{MN} = \og_{MN} + g^{(>)}_{MN} \, , \quad\quad
g^{(>)}_{MN} \equiv \sum_{n>0} g^{(n)}_{MN} \, , 
} 
order by order in an expansion in time ($x^+$) derivatives.  We will
say that $g^{(n)}$ ``has $d_t = n$'' in this expansion.  This is a
nontrivial solution since it corresponds to a nontrivial ``tangent
deformation'' in the moduli space of static solutions.  The
corrections $g^{(n)}$ vanish on the spatial infinity.

 The calculation is set up in Section 2, where we recall the uplifting
of the static solution to 11d, and discuss the solution for moving
sources to lowest order in transverse velocities.  In section 3 we
determine the leading large distance behaviour of the solution for up
to three centers.  In section 4 we give the result of the calculation
for the leading order two and three particle scattering.

 In the following $V$ will stand for a typical (small) transverse speed
and $L$ a typical (large) transverse separation.

\newsec{Uplifting to 11d}

 The static BPS solution of IIA supergravity with $n_c$ D0-branes (\dob,
and references therein) is,
\eqn\jmtwa{
\eqalign{
d{\overline{s}}^2 &= \og_{\mu\nu}(x)dx^\mu dx^\nu 
= - f_0(\vy)^{-1/2} dt^2 + f_0(\vy)^{1/2} d\vy \cdot d\vy \cr
e^{\oPhi(x)} &= g_s f_0^{3/4} \, , \quad\quad \oA_0(x) = f_0^{-1} - 1 \cr
f_0(\vy) &= 1 + \sum_{i=1}^{n_c}f_0^{(i)} \, , \quad\quad 
f_0^{(i)} = {\mu_i \over {|\vy - \vy_i|^7}} \, . \cr
}
}
The statement that these are D0-branes
means that the ``charges'' in this solution, $\mu_i$, are determined
in terms of string parameters \polchrr\ by $\mu_i \sim \ell_s^7 g_s$.
Using the Kaluza-Klein relation we can lift this to a solution of 11d
supergravity with the 11th direction compactified on a spacelike
circle of radius $R_c$, with vanishing 3-form, and
\eqn\jmtwb{
\eqalign{
ds^2 &= \og_{MN} (x) dx^M dx^N = \oD^{-1/8} \og_{\mu\nu}dx^\mu dx^\nu +
\oD (dx^{11} + \oA_\mu dx^\mu)^2 \, , \cr
&= (f_0(\vy) - 2) dt^2 + f_0(\vy) (dx^{11})^2 -2 (f_0(\vy) - 1) dt dx^{11} +
d\vy\cdot d\vy\, ,\cr 
} 
} 
where $\oD = e^{4 \oPhi/3} = f_0$.  In lightcone coordinates, 
$x^\pm = x^{11} \pm t$, the result is suggestively simple \hyunetal,
\eqn\jmtwc{
\eqalign{
ds^2 &= dx^+ dx^- + f_0^B(\vy) dx^- dx^- + d\vy \cdot d\vy  \cr
f_0^B(\vy) &= f_0(\vy) - 1 
= \sum_{i=1}^{n_c}{\mu_i \over {|\vy - \vy_i|^7}} \, . \cr
}
}

 The corresponding D0-brane source in 11d supergravity is (in the
leading approximation where we neglect spin effects) a massless scalar
with fixed nonzero momentum in the compact direction.  The appropriate
point particle action has been discussed in \bbpt.  The point
particle action in 11d for a {\it massive} ``particle $i$'' with mass
$m_i$, is
\eqn\jmtwe{
S^{(i)}_m = - m \int{d\tau} \big[- g_{MN} {d x^M \over 
{d\tau}}{d x^N \over {d\tau}} \big]^{1/2} \, .
}
Choosing to parametrize the trajectory by $x^+$, this gives
\eqn\jmtwf{
p^i_- = m {g_{-+} + g_{-a} v_i^a + g_{--} v_i^- \over 
{[- g_{++} - g_{+a} v_i^a - g_{ab} v_i^a v_i^b - 
2 g_{+-} v_i^-  - 2 g_{-a} v_i^- v_i^a  - g_{--} v_i^- v_i^- ]^{1/2}}} \, ,
}
where $v^M_i = dy^M_i/d x^+$.  Assuming no $x^-$ dependence, then
$p^i_-$ is a cyclic variable, and the corresponding ``Routhian'' for
the constant $p^i_-$ physics of the {\it massless} particle is
$$
S^{(i)} = \lim_{m \rightarrow 0} (S^{(i)}_m - \int{dx^+}p^i_-  v_i^-) \, .
$$
To implement this limit is very easy, since the first term vanishes
and $v_i^-$ is determined at lowest order by the vanishing of the
denominator in \jmtwf.  Thus, with $p^i_- = Q_i$, the D0-brane source is
\eqn\jmtwg{
S^{(i)} = Q_i \int {dx^+}{1\over {g_{--}}} \big[ g_{+-} + g_{-a} v_i^a -
\sqrt{(g_{+-} + g_{-a} v_i^a)^2 - g_{--} (g_{++} + 2 g_{+a} v_i^a + 
g_{ab} v_i^a v_i^b)} \big] \, .
}

 In summary, then, the system of interest is described by
\eqn\jmtwm{
S = S_E + \sum_{i=1}^{n_c} S^{(i)} \, , 
} 
where $S_E = \kappa^{-2}\int{d^{11}x}\sqrt{-g}R(g)$ and $S^{(i)}$ is
specified in \jmtwg.  We work in the lightcone frame with $x^- \sim
x^- + 2\pi R$.  We may now proceed with the $d_t$-expansion of the
$x^-$-independent solution for centers moving with slow transverse
velocity, and then substitute in to determine the effective action for
the centers.

 To zeroth order we simply have the static solution which fixes
\eqn\jmtwl{
Q_i = {32 \pi^4 \over {15}} \mu_i {2 \pi R \over {\kappa^2}} \, .
}
To first order the solution is easily understood in terms
of transverse boosts.  A ``Galilean'' transversal boost with velocity
$\vv$ in lightcone time, as is appropriate for our discussion, gives
\eqn\jmtwh{
\eqalign{
x'^+ &= x^+ \cr
x'^- &= x^- + 2 \vv \cdot \vy - v^2 x^+ \cr
\vy' &= \vy - x^+ \vv \, . \cr 
} 
} 
Taking $x'$ as the coordinates of the ``static frame'', we obtain the
metric for a single center moving transversally with constant velocity
by using the boost as a coordinate transformation,
\eqn\jmtwi{
\eqalign{
ds^2 =& (1 - 2 v^2 f_0^B(r')) dx^+ dx^- + f_0^B(r') dx^- dx^- + 
v^4 f_0^B (r') dx^+ dx^+ + \cr
& + (\delta_{ab}  + 4 f_0^B(r') v_a v_b) dy^a dy^b 
 + 4 f_0^B(r') v_a dx^- dy^a - 4 f_0^B(r') v^2 v_a dx^+ dy^a \, , 
\cr
} 
}
where $r' = |\vy - x^+ \vv|$.  This is extended to $n_c$ centers
moving independently to give the ansatz 
\eqn\jmtwj{
\eqalign{
d\hs^2 =& \left(1 - 2 \sum_{i=1}^{n_c} v_i^2 f_0^{(i)}(r_i)\right) dx^+ dx^- + 
\sum_{i=1}^{n_c}f_0^{(i)}(r_i) dx^- dx^- + 
\sum_{i=1}^{n_c}v_i^4 f_0^{(i)}(r_i) dx^+ dx^+ + \cr
 & + \left(\delta_{ab}  + 
4 \sum_{i=1}^{n_c}f_0^{(i)}(r_i) v_i^a v_i^b\right) dy^a dy^b 
 + 4 \sum_{i=1}^{n_c}f_0^{(i)}(r_i) v_i^a dx^- dy^a - \cr
& - 4 \sum_{i=1}^{n_c}f_0^{(i)}(r_i) v_i^2 v_i^a dx^+ dy^a \, , 
\cr
} 
} 
where $r_i = |\vy - \vy_i(x^+)|$.  We will denote by $\hg_{MN}$ the
metric specified by \jmtwj.  In this way we find the possible nonzero
first order corrections to the metric are just those which implement
the ``constraints'' of the lightcone parametrization,
\eqn\jmtwk{
\eqalign{
g^{(1)}_{+a} &= \hg^{(1)}_{+a} = 0 \cr 
g^{(1)}_{-a} &= \hg^{(1)}_{-a} = \sum_i 2 {\mu_i \over {r_i^7}}
v_i^a \, . \cr 
} 
} 
It is now straightforward to check that the independently boosted
metric $\hg$ is a solution to first order in $V$ of the 11d system 
\jmtwm.

\newsec{Leading large-distance behaviour of the $d_t$ expansion}

 In the $d_t$-expansion we must iteratively determine the expression
for $\dot{\vv}_i$, which clearly only receives corrections at even
orders of $d_t$.  It is well known that the $d_t = 2$ contribution
vanishes (see, e.g., \flato\flattw) -- this is ``flatness of the moduli
space''.  To see this in the present calculation we simply evaluate
the effective action to second order, for which we only need the first
order solution given previously.  The resulting effective Lagrangian
is a total derivative.  Equivalently, note that from the equation of
motion for the centers (the geodesic equation), using \jmtwk\ we have
(to second order)
\eqn\jmtha{
\dot{v}_i^a = - {1\over 2} \partial_a g^{(2)}_{++} + O(V^4) \, .
}
But from the Einstein equations to second order, we find
\eqn\jmthb{
- {\pi R\over \kappa^2} \Delta_\perp g^{(2)}_{++} = T^{(2)}_{++} = 0 \, ,
}
and thus $\dot{\vv}_i \sim O(V^4)$ as stated.  

 Further, a detailed calculation\foot{To regularize the point particle
we replace $r_i \rightarrow (r_i^2 + \epsilon^2)^{1/2}$.} \aj\ shows
that
\eqn\jmthc{
g^{(2)}_{ab} = \hg_{ab} \, , \quad\quad g^{(2)}_{+-} = \hg_{+-} \, ,
}
and 
\eqn\jmthd{
g^{(2)}_{--} = \sum_{ij} \mu_i\mu_j {|\vv_i - \vv_j|^2 \over 
{|r_i|^7 |r_j|^7}} + f^{(2)} \, ,
}
where,
$$
\Delta_\perp f^{(2)} = 2 \partial_a\partial_b \sum_{ij} \mu_i\mu_j 
{(v_i - v_j)^a (v_i - v_j)^b \over {|r_i|^7 |r_j|^7}} \, .
$$
The important result in the above is the observation that the solution
at second order differs from the boosted metric \jmtwi, \jmtwj\ by
$O({V^2 \over {L^{14}}})$.  

 At higher order the leading behaviour is equally simple.  Let us
separate out the ``boosted'' corrections $\oh$ which are contained in
\jmtwi\ and \jmtwj; ie,
\eqn\jmthe{
g^{(>)}_{MN} = \oh_{MN} + h_{MN} \, ,
}
where $\hg_{MN} = \og_{MN} + \oh_{MN}$.  Then
\eqn\jmthf{
h^{(n)}_{MN} = O\left({V^n \over {L^{14}}}\right) \, .
}
To see this one just calculates the leading terms in the Einstein
equations at $n$th order in the expansion.  For the Einstein tensor
these are terms involving $g^{(n)}_{MN}$, since products of
lower order (in $d_t$) contributions will be higher order in $1/L$.
The $1/L$ expansion of the source is similarly straightforward,
and the leading terms are just those required for $\oh^{(n)}_{MN}$.  

 We now show that the leading term in the effective action is
determined by $\hg_{MN}$.  Expanding the action around $\hg_{MN}$,
the above result implies that
\eqn\jmthg{
S[g] = S[\hg] + {\delta S \over {\delta g}}\big|_{\hg}[h] + 
{\rm higher \, order} \, .
}
The second term can be further expanded around $\og_{MN}$, and
only the first term in this expansion is required for the leading order
result,
$$
{\delta S \over {\delta g}}\big|_{\og + \oh}[h] = 
{\delta^2 S \over {\delta g \delta g}}\big|_{\og}[\oh,h] + 
{\rm higher \, order}\, .
$$
Now, the fact that the boosted single center metric is a solution of
the Einstein equations for a constant transverse velocity source
implies that, up to terms with derivatives on $\vv$,
$$
{\delta S \over {\delta g}}\big|_{\hg_i}[h] = 0 \, ,
$$
where $\hg_i$ denotes the boosted single center solution (for
the $i$th center) -- ie, the limit of $\hg_{MN}$ as $\mu_j \rightarrow 0$,
$j\neq i$.  Thus we must have, up to terms with derivatives on $\vv$,
$$
{\delta^2 S \over {\delta g \delta g}}\big|_{\og}[\oh,\cdot] =
O({\mu_i \mu_j \over {L^{14}}}) \, ,
$$
meaning that the RHS is at least quadratic in the $\mu_i$ since it
vanishes if all but one of them is sent to zero.  But then this is higher
order, and can be ignored.  Thus we only have to worry about the
contribution of $\dot{\vv}$ terms ($\ddot{\vv}$ terms are 
irrelevant, as is easily seen by the following argument).

 Using the previous results, we have so far shown that, in the second
term of \jmthg, 
\eqn\jmthh{ 
{\delta S \over {\delta g}}\big|_{\og + \oh} = O(\mu V^4) \, .  
}
Thus, we only need the $\dot{\vv}$ terms in the LHS of \jmthh, and
they need only be contracted with $h^{(2)}$ to this order.  A
calculation shows that in the Einstein tensor, $\dot{\vv}$ terms only
appear, at this order, in $G_{+-}$ and $G_{ab}$.  But, as summarized
above, $h^{(2)}_{+-}$ and $h^{(2)}_{ab}$ vanish.  Further, any 
contractions with off-diagonal metric components are higher order.
Thus, finally, the result is proved -- all terms but the first in \jmthg\
are higher order in $1/L$.

\newsec{Computing the action}

 At this point we simply compute the leading contribution up to
$O(V^6)$.  The result for the leading $O(V^4)$ contribution to two
particle scattering is ( we don't write the ``polarization'' terms,
with numerators containing $\vv \cdot \vy$)
\eqn\jmfoa{
S_{\rm eff}^{(4)} = - {32 \pi^4 \over {15}} {2\pi R\over {\kappa^2}} 
\mu_1\mu_2 {|\vv_1 - \vv_2|^4 \over {|\vy_1 - \vy_2|^7}} \, .  
} 
This is precisely the result reported in \beco.  In the present
calculation it results from a cancellation between Einstein and source
contributions.

 The result for the leading $O(V^6)$ contribution to three particle
scattering, in the limit considered by Dine and Rajaraman ($|\vy_3| >>
|\vy_1 - \vy_2|$) is (for brevity we only write the ``Dine-Rajaraman'' term)
\eqn\jmfoa{ 
S_{\rm eff}^{(6)} = -4 {32 \pi^4 \over {15}} {2\pi R\over {\kappa^2}}
\mu_1\mu_2\mu_3 {|\vv_1 - \vv_2|^2 |\vv_2 - \vv_3|^2 |\vv_1 - \vv_3|^2
\over {|\vy_3|^7 |\vy_1 - \vy_2|^7}} \, .  
}  
It is interesting to note that there is clearly no contribution from
the source action of this form.  

 Hence we see that the term required for agreement with the
perturbative supergravity calculation does appear.  The technical
calculation in this paper is simply a check of the standard
IIA/M-theory relation.  The significance to Matrix theory rests on the
conjectured relation to the large $N$ limit, which it is clearly of
interest to understand better.

 After this paper was finished a number of papers have appeared which
discuss the Dine-Rajaraman problem \ffi\wtr\ecgr\okyo\ from the finite
$N$ side.  In \ffi\ it was suggested that the supersymmetry
cancellations proposed in the Matrix theory calculation of \dira\
would not occur, but this has been disputed in \wtr\ and \ecgr.  The
technical calculation in the present paper has significant overlap
with \okyo, where, further, the Matrix theory result is recalculated
and shown to be in agreement at finite $N$.  The present paper
supports the supergravity side of this calculation.

\listrefs

\vfil\eject
\end